\documentclass[twocolumn,prl,floatfix,showpacs,preprintnumbers,nofootinbib,%
superscriptaddress]{revtex4-2}
\usepackage{mathrsfs}
\usepackage{amssymb}	
\usepackage{amsmath}
\usepackage{graphicx}
\usepackage{subdepth}
\usepackage{xcolor}
\definecolor{lcolor}{rgb}{0.5,0,0}
\definecolor{citcolor}{rgb}{0,0.3,0.0}

\usepackage[breaklinks,colorlinks,urlcolor=blue,citecolor=citcolor,linkcolor=lcolor]{hyperref}

\usepackage{tikz}

\newcommand{\Pt}{{\mathbf{P}}}

\newcommand{\xt}{{\mathbf{x}}}

\newcommand{\kt}{{\mathbf{k}}}

\newcommand{\epst}{\boldsymbol{\varepsilon}}

\newcommand{\kaz}{\kappa_{z}}
\newcommand{\kac}{\kappa_{\chi}}

\newcommand{\nlo}{{\textnormal{NLO}}}

\newcommand{\epsl}{{\varepsilon\!\!\!/}}

\newcommand{\ud}{\, \mathrm{d}}

\newcommand{\nc}{{N_\mathrm{c}}}

\newcommand{\cf}{C_\mathrm{F}}

\newcommand{\nr}[1]{(\ref{#1})}

\newcommand{\as}{\alpha_{\mathrm{s}}}

\newcommand{\fig}{Fig.~}
\newcommand{\figs}{Figs.~}
\newcommand{\eq}{Eq.~}

\newcommand{\eqs}{Eqs.~}

\newcounter{diag}
\newcommand{\labeldiag}[1]{\refstepcounter{diag} \label{#1}}

\begin{document}

\author{G. Beuf}
\affiliation{
Theoretical Physics Division, National Centre for Nuclear Research,
Pasteura 7, Warsaw 02-093, Poland
}

 \author{T. Lappi}
\affiliation{
Department of Physics, %
 P.O. Box 35, 40014 University of Jyv\"askyl\"a, Finland}
\affiliation{
Helsinki Institute of Physics, P.O. Box 64, 00014 University of Helsinki,
Finland}

\author{R. Paatelainen}
\affiliation{
Department of Physics, P.O. Box 64, 00014 University of Helsinki,
Finland}
\affiliation{
Helsinki Institute of Physics, P.O. Box 64, 00014 University of Helsinki,
Finland}

\title{Massive quarks at one loop in the dipole picture of Deep Inelastic Scattering}
\preprint{HIP-2021-46/TH}


\begin{abstract}
We calculate the light cone wave functions for a virtual photon to split into quark-antiquark states, including for the first time quark masses at one loop accuracy. These wave functions can be used to calculate cross sections for several precision probes of perturbative gluon saturation at the Electron-Ion Collider. Using these wave functions we derive, for the first time, the  
dipole picture DIS cross sections 
at one loop 
for longitudinal and transverse virtual photons  
including
quark masses. The quark masses are renormalized in the pole mass scheme, satisfying constraints from  the requirement of Lorentz invariance of the quark Dirac and Pauli form factors. 
\end{abstract}

\maketitle

\section{Introduction}

It is believed that in very high energy hadronic collisions, the partonic constituents of hadrons and nuclei exhibit a qualitatively new kind of gluon saturation behavior, characterized by strong nonlinear interactions even at short distance scales where the coupling is weak. An experimentally clean way to study this regime are high energy Deep Inelastic Scattering (DIS) experiments. Studying  gluon saturation  is a key science goal of the  future Electron-Ion Collider (EIC)~\cite{Accardi:2012qut,AbdulKhalek:2021gbh}, which will address it with a broad program of precision measurements. The EIC can reach further into the saturation regime than previous measurements at HERA, because it also collides heavy nuclei, where saturation phenomena are enhanced~\cite{Kowalski:2007rw}.

One could search for signals of gluon saturation in the renormalization group evolution of cross sections as functions of the kinematical variables  $Q^2$ and $x$~\cite{Albacete:2012rx,Marquet:2017bga}.  With the EIC collision energy, however, the kinematical leverarm to distinguish fine details or asymptotic features of evolution is limited, since evolution is only logarithmic in  $Q^2$ or $x$. Instead, one must most likely look for evidence of saturation in a combination of high precision measurements of different processes. 
A key part in such a comprehensive set of studies is played by heavy quark observables, both inclusive and exclusive. Of particular interest are processes involving charm quarks, where the quark mass is heavy enough to justify a weak coupling treatment, but light enough to be sensitive to saturation effects. In a collinear factorization picture, the charm cross section is one of the most sensitive probes of small-$x$ gluons at the EIC~\cite{Aschenauer:2017oxs}. To access gluon saturation it is better to use instead the coordinate space dipole picture \cite{Nikolaev:1990ja,Kopeliovich:1991pu,Nikolaev:1991et,Mueller:1993rr,Mueller:1994jq,Mueller:1994gb} of DIS, where the virtual photon emitted by the electron first splits to partonic constituents, which then eikonally interact with the target. The dipole picture naturally involves the eikonal scattering amplitudes, Wilson lines, used to quantify gluon saturation in the CGC picture~\cite{McLerran:1993ni,Weigert:2005us,Gelis:2010nm}. In the dipole picture light quarks  are affected by contributions of nonperturbatively large dipoles in the ``aligned jet'' configurations~\cite{Mantysaari:2018nng,Mantysaari:2018zdd}, but the mass of the heavy quark makes them safe from this problematic part of phase space.

The theoretical framework of choice to understand saturation and the dipole picture in high energy DIS is  QCD light cone perturbation theory~\cite{Kogut:1969xa,Bjorken:1970ah,Lepage:1980fj,Brodsky:1997de} (LCPT). Here one first calculates the photon \emph{light cone wave function} (LCWF) describing the probability amplitude of the photon to split into a partonic state. The LCWF is a universal quantity in perturbative field theory. It is a necessary ingredient in cross section calculations for different  inclusive and  exclusive scattering processes~\cite{Kowalski:2006hc,Watt:2007nr,Rezaeian:2012ji,Mantysaari:2018zdd,Boussarie:2014lxa,Boussarie:2016bkq,Boussarie:2016ogo,Lappi:2020ufv,Escobedo:2019bxn,Mantysaari:2021ryb,Mantysaari:2022bsp}. Recently, the photon LCWF has been calculated to one loop accuracy in QCD perturbation theory~\cite{Beuf:2016wdz,Beuf:2017bpd,Hanninen:2017ddy} leading to a description of the HERA inclusive cross section~\cite{Beuf:2020dxl} with massless quarks (see also \cite{Balitsky:2010ze,Balitsky:2012bs}). In this letter we report the result of the calculation of the so far unknown  NLO $\gamma^*\to q\bar{q}$ wavefunction with massive quarks.

This Letter will be accompanied by a longer paper \cite{Beuf:2022ndu} with full technical details on the calculation for the transverse photon,  the longitudinal photon having  already been presented in Ref~\cite{Beuf:2021qqa} (see also~\cite{Dai:2022imf}). In a separate follow-up paper we will discuss the issue of quark mass renormalization in LCPT in more detail. 

\begin{figure}[tbh]
\label{diag:lo}
\centerline{
\resizebox{0.30\textwidth}{!}{
\mbox{
\includegraphics[width=5.0cm]{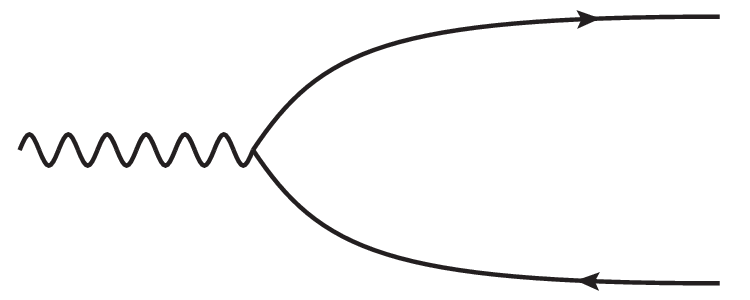}
\begin{tikzpicture}[overlay]
\draw [dashed] (-2,2.1) -- (-2,-0.1);
\node[anchor=west] at (-1.3,1.6) {\large $k_0^+, \ \kt_0, \ h_0$};
\node[anchor=west] at (-1.3,0.5) {\large $k_1^+, \ \kt_1, \ h_1$};
\node[anchor=north] at (-5,0.9) {\large $\gamma_{L,T},\  q^\mu$};
\end{tikzpicture}
}
}
}
\caption{The only diagram for the virtual photon-to-quark-antiquark wavefunction at leading order. There is one energy denominator, denoted with a dashed line.}
\label{fig:lovertex}
 \end{figure}
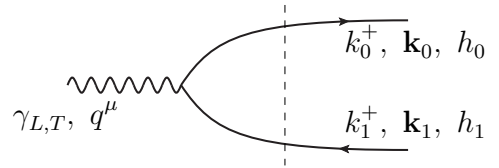

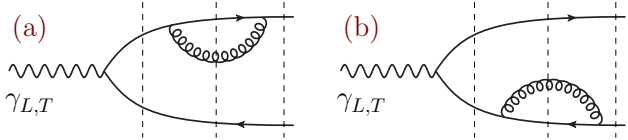
\begin{figure}[tbh]
\labeldiag{diag:oneloopSEUP}
\labeldiag{diag:oneloopSEDOWN}
\centerline{
\resizebox{0.24\textwidth}{!}{
\includegraphics[width=5.4cm]{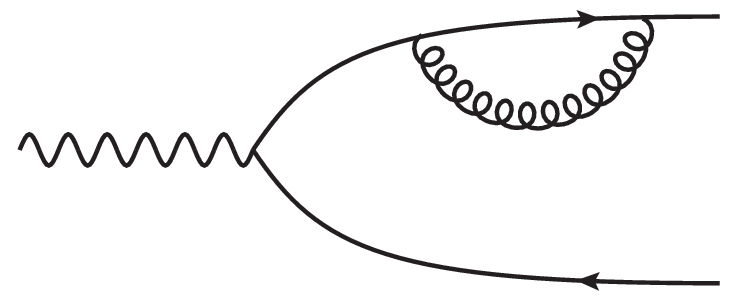}
\begin{tikzpicture}[overlay]
\draw [dashed] (-3,2.3) -- (-3,-0.1);
\draw [dashed] (-1.7,2.3) -- (-1.7,-0.1);
\draw [dashed] (-0.5,2.3) -- (-0.5,-0.1);
\node[anchor=north] at (-5,0.8) {\Large $ \gamma_{L,T}$};
\node[anchor=north] at (-5,2.2) {\Large\ref{diag:oneloopSEUP}};
\end{tikzpicture}
}
\resizebox{0.24\textwidth}{!}{
\includegraphics[width=5.4cm]{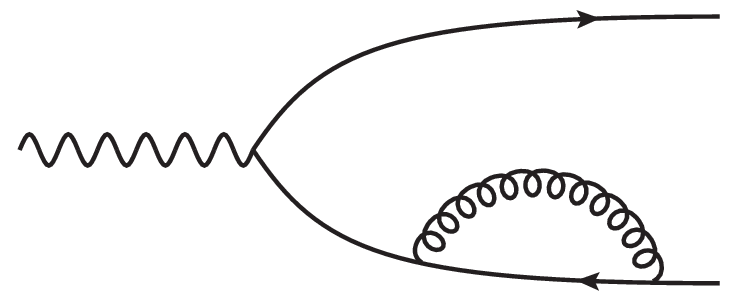}
\begin{tikzpicture}[overlay]
\draw [dashed] (-3,2.3) -- (-3,-0.1);
\draw [dashed] (-1.7,2.3) -- (-1.7,-0.1);
\draw [dashed] (-0.5,2.3) -- (-0.5,-0.1);
\node[anchor=north] at (-5,0.8) {\Large $\gamma_{L,T}$};
\node[anchor=north] at (-5,2.2) {\Large\ref{diag:oneloopSEDOWN}};
\end{tikzpicture}
}
}
\caption{Quark self-energy diagrams for the $\gamma^*\to q\bar{q}$ LCWF, with 3 energy denominators (dashed lines)} 
\label{fig:selfenergy}
 \end{figure}

\begin{figure}[tbh]
\labeldiag{diag:oneloopSEUPinst}
\labeldiag{diag:oneloopSEDOWNinst}
\centerline{
\resizebox{0.24\textwidth}{!}{
\includegraphics[width=5.4cm]{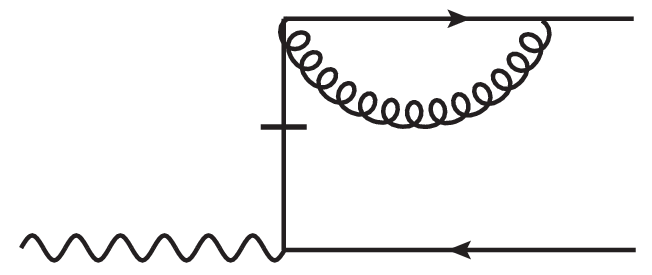}
\begin{tikzpicture}[overlay]
\draw [dashed] (-2,2.3) -- (-2,-0.1);
\draw [dashed] (-0.6,2.3) -- (-0.6,-0.1);
\node[anchor=south] at (-4.8,0.4) {\Large $\gamma_{T}$};
\node[anchor=north] at (-5,2.2) {\Large\ref{diag:oneloopSEUPinst}};
 \end{tikzpicture}
}
\resizebox{0.24\textwidth}{!}{
\includegraphics[width=5.4cm]{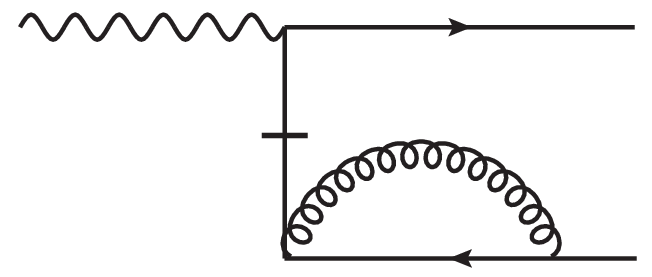}
\begin{tikzpicture}[overlay]
\draw [dashed] (-2,2.3) -- (-2,-0.1);
\draw [dashed] (-0.6,2.3) -- (-0.6,-0.1);
\node[anchor=north] at (-4.8,1.9) {\Large $\gamma_{T}$};
\node[anchor=north] at (-5,0.8) {\Large\ref{diag:oneloopSEDOWNinst}};
 \end{tikzpicture}
 }
}
\caption{Instantaneous self-energy diagrams for the $\gamma^*\to q\bar{q}$ LCWF, with 2 energy denominators (dashed lines); these do not appear for longitudinal photons. 
} 
\label{fig:selfenergyinst}
 \end{figure}
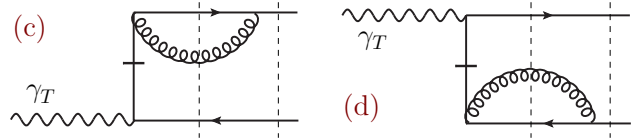

 \begin{figure}[t!]
\labeldiag{diag:vertexqbarem}
\labeldiag{diag:vertexqem}
\centerline{
\resizebox{0.24\textwidth}{!}{
\includegraphics[width=5.4cm]{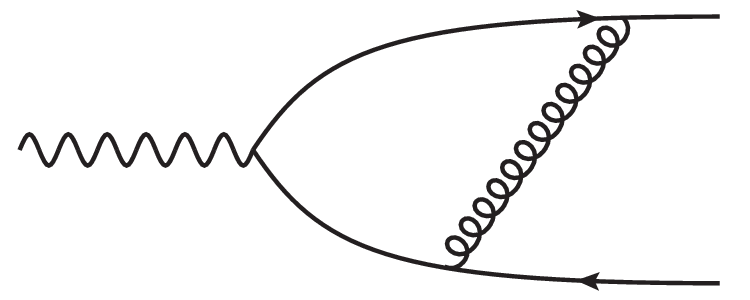}
\begin{tikzpicture}[overlay]
\draw [dashed] (-3,2.3) -- (-3,-0.1);
\draw [dashed] (-1.7,2.3) -- (-1.7,-0.1);
\draw [dashed] (-0.6,2.3) -- (-0.6,-0.1);
\node[anchor=north] at (-5,0.8) {\Large $\gamma_{L,T}$};
\node[anchor=north] at (-5,2.2) {\Large\ref{diag:vertexqbarem}};
 \end{tikzpicture}
}
\resizebox{0.24\textwidth}{!}{
\includegraphics[width=5.4cm]{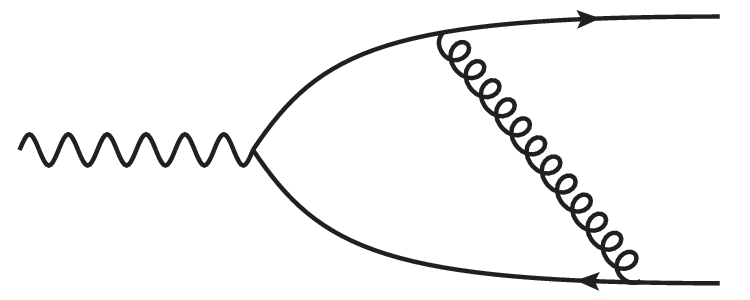}
\begin{tikzpicture}[overlay]
\draw [dashed] (-3,2.3) -- (-3,-0.1);
\draw [dashed] (-1.7,2.3) -- (-1.7,-0.1);
\draw [dashed] (-0.6,2.3) -- (-0.6,-0.1);
\node[anchor=north] at (-5,0.8) {\Large $\gamma_{L,T}$};
\node[anchor=north] at (-5,2.2) {\Large\ref{diag:vertexqem}};
 \end{tikzpicture}
}
}
\caption{Vertex correction diagrams, with 3 energy denominators.}
\label{fig:vertexT}
\end{figure}
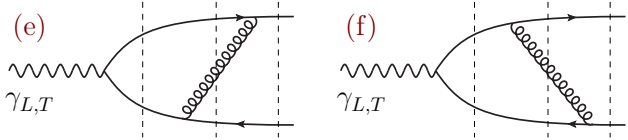

\begin{figure}[tbh]
\labeldiag{diag:oneloopvertexinst1}
\labeldiag{diag:oneloopvertexinst2}
\labeldiag{diag:gluoninst}
\centerline{
\resizebox{0.17\textwidth}{!}{
\includegraphics[width=3.5cm]{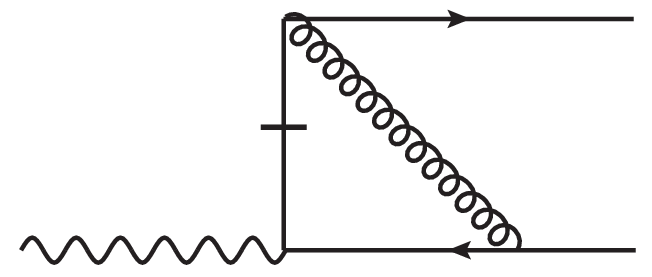}
\begin{tikzpicture}[overlay]
\draw [dashed] (-1.467,1.5333) -- (-1.467,-0.06667);
\draw [dashed] (-0.4,1.5333) -- (-0.4,-0.06667);
\node[anchor=south] at (-3.2,0.15) {\Large $\gamma_{T}$};
\node[anchor=north] at (-3.3333,1.467) {\large\ref{diag:oneloopvertexinst1}};
 \end{tikzpicture}
}
\resizebox{0.17\textwidth}{!}{
\includegraphics[width=3.5cm]{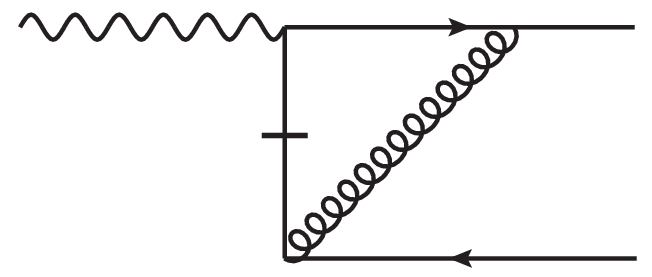}
\begin{tikzpicture}[overlay]
\draw [dashed] (-1.467,1.5333) -- (-1.467,-0.06667);
\draw [dashed] (-0.4,1.5333) -- (-0.4,-0.06667);
\node[anchor=north] at (-3.2,1.1) {\Large $\gamma_{T}$};
\node[anchor=north] at (-3.33333,0.5333) {\large\ref{diag:oneloopvertexinst2}};
 \end{tikzpicture}
}
\resizebox{0.17\textwidth}{!}{
\includegraphics[width=3.5cm]{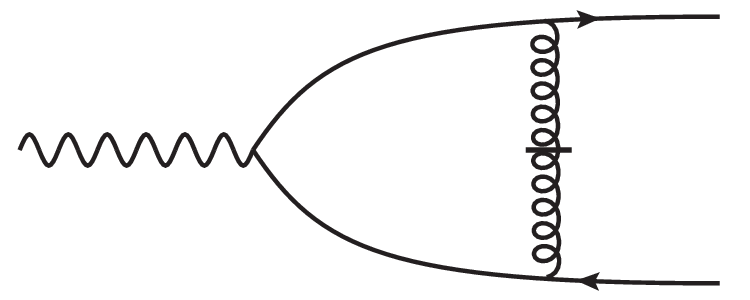}
\begin{tikzpicture}[overlay]
\draw [dashed] (-1.467,1.5333) -- (-1.467,-0.06667);
\draw [dashed] (-0.4,1.5333) -- (-0.4,-0.06667);
\node[anchor=north] at (-3.2,0.53333) {\Large $\gamma_{L,T}$};
\node[anchor=north] at (-3.333,1.4) {\large\ref{diag:gluoninst}};
 \end{tikzpicture}
}
}
\caption{Instantaneous vertex correction diagrams, only the last one appears for longitudinal photons. } 
\label{fig:instvertex}
 \end{figure}
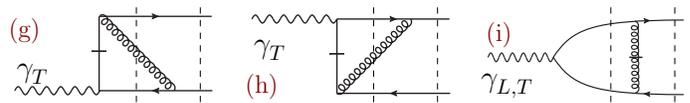

\section{Calculational setup}

\paragraph{Loop calculations in LCPT}
We use the Hamiltonian LCPT formulation of perturbative QCD~\cite{Kogut:1969xa,Bjorken:1970ah,Lepage:1980fj,Brodsky:1997de}. This approach is an ideal one for high energy scattering, where particles move on lightlike trajectories and their dynamics is thus naturally quantized with a lightlike time coordinate. The additional advantage of the LCPT formulation in light cone gauge is that only physical degrees of freedom are present in the calculation, which comes at the expense of having additional ``instantaneous'' 4-particle interactions resulting from the exchanges of non-propagating degrees of freedom.  An unfortunate disadvantage is that because of the separate treatment of longitudinal and transverse coordinates, the theory is not manifestly Lorentz-invariant at the quantum level. 

In the LCPT approach, one develops the full quantum state of the incoming particle, in this case the virtual photon, in a Fock state expansion of bare states. At  small-$x$, the partons interact with the color fields of the target, thus only Fock states consisting of quarks and gluons are of relevance here. The leading such component in the photon state is the quark-antiquark dipole, depicted in \fig\ref{fig:lovertex}. At NLO one also needs to include corrections from gluon loops, and gluon emission diagrams, i.e. $q\bar{q}g$ Fock states. 

The coefficients of the expansion of the interacting (photon) state in terms of bare states are known as light cone wave functions. Perturbatively they are obtained in terms of a set of diagrammatical rules~\cite{Brodsky:1997de,Kovchegov:2012mbw}. For every vertex one includes a matrix element depending on the helicities, polarizations and momenta of the participating particles. Instantaneous vertices are denoted by vertical crossed lines (time propagates from left to right). For every intermediate state (including the final state), one includes a light cone energy denominator which, in a covariant perturbation theory language, originates from integrating over the light cone energy $k^-$  and setting it on shell using the pole of a propagator. One then integrates over loop momenta and sums over internal helicities. 

The leading order $\gamma^*\to q\bar{q}$ wavefunction (see e.g. Refs.~\cite{Nikolaev:1990ja,Nikolaev:1991et,Mueller:1993rr,Kowalski:2006hc}) is obtained by evaluating the diagram of \fig\ref{fig:lovertex}, with one gauge boson-fermion vertex and one energy denominator. The diagrams needed for the NLO calculation are the same as in the massless case~\cite{Beuf:2016wdz,Beuf:2017bpd,Hanninen:2017ddy}, as are most other calculational details. For our calculation we need first the self-energy corrections for the fermions, \fig\ref{fig:selfenergy}. For transverse photons, there is also a self-energy correction from an instantaneous interaction, \fig\ref{fig:selfenergyinst}. Longitudinal photons are properly speaking parts of an instantaneous interaction with the lepton~\cite{Beuf:2011xd}, and thus do not have instantaneous vertices. 
The photon-quark-antiquark vertex gets corrections from normal physical gluons, \fig\ref{fig:vertexT}, and also from instantaneous interactions, \fig\ref{fig:instvertex}. Gluon emission can happen via normal, \fig\ref{fig:qgqbar}, and, again only for transverse photons, instantaneous interactions, \fig\ref{fig:qgqbarinst}. We have evaluated all these diagrams.

The loop momenta are integrated over in $2-2\varepsilon$ transverse dimensions, with a cutoff $\alpha$ regularizing any soft divergences arising from longitudinal momentum integrals in the $k^+\to 0$ limit.  After integrating over the loop momenta one sums over internal helicities and gluon polarizations.  We have performed the helicity sums  both in the conventional dimensional regularization (CDR) scheme as in Ref.~\cite{Beuf:2016wdz,Beuf:2017bpd}, and in the four-dimensional helicity (FDH) scheme as in Ref.~\cite{Hanninen:2017ddy}, with equal results for the cross sections.

\paragraph{Issue of mass renormalization }
At this order in perturbation theory also  the quark mass is renormalized. We work in the  Hamiltonian LCPT approach, where from the Lorentz-invariant Lagrangian one first derives  a Hamiltonian, and only then sets out to canonically quantize the latter in light cone gauge $A^+=0$. In the Hamiltonian the fermion mass appears in two separate terms~\cite{Burkardt:1991tj}.  The free part of the Hamiltonian has a ``kinetic mass'', which determines the relation between light cone energy $k^-= (\kt^2+m^2)/(2k^+)$ and 3-momentum $(\kt,k^+)$. There is also the ``vertex mass,'' the coefficient of the light cone helicity flip term of the gauge boson-$q\bar{q}$ vertex (see \eq\nr{eq:melem}). The latter did not need to be renormalized in  our earlier calculation~\cite{Beuf:2021qqa} of the longitudinal photon state.

Lorentz-invariance at the original Lagrangian level guarantees that the kinetic and vertex masses are equal in nature. Regularization methods that break Lorentz-invariance, such as the  transverse dimensional regularization combined with longitudinal cutoffs used in our previous calculations for massless quarks Refs.~\cite{Beuf:2016wdz,Beuf:2017bpd,Hanninen:2017ddy}, require the restoration of this invariance at the loop level by separate renormalization conditions for the kinetic and vertex masses. This was already known from the pioneering LCPT calculations of Refs.~\cite{Mustaki:1990im,Zhang:1993is,Zhang:1993dd,Harindranath:1993de}. One can, however, slightly modify the regularization procedure by including, in addition to the diagrams appearing here, also the  ``self-induced inertia'' or ``seagull'' diagrams~\cite{Pauli:1985pv,Tang:1991rc,Brodsky:1997de} before the integrations. In the latter case, it becomes possible to maintain the equality of the vertex and kinetic masses.

\paragraph{Spinor structure }
Our calculation is organized in terms of  possible independent spinor structures of the wave function. For a transversally polarized photon at leading order,  the spinor structure of the leading order light-cone gauge $\gamma^*(q)\to q(k_0)\bar{q}(k_1)$ matrix element can be decomposed (see e.g.~\cite{Beuf:2021qqa}) in terms of three independent spinor structures as
\begin{multline} \label{eq:melem}
\bar{u}(0)\epsl_{\lambda}(q) v(1) = 
\frac{q^+}{2k_0^+k_1^+} 
\Biggl \{ 
\biggl[
\frac{k_0^+-k_1^+}{q^+} 
\delta^{ij}
\bar{u}(0)\gamma^{+}v(1)  
\\
+ \frac{1}{2}\bar{u}(0)\gamma^{+}[\gamma^i, \gamma^j]v(1)  \biggr] \Pt^{i}  
- m\bar{u}(0)\gamma^{+}\gamma^{j}v(1)  \Biggr \} \epst^{j}_{\lambda},
\end{multline}
where $i,j$ are transverse indices and $\Pt = (k_1^+/q^+)\kt_0 - (k_0^+/q^+)\kt_1$ is  the $q\bar{q}$ relative transverse momentum. The result for the $\gamma^* \to q\bar{q}$ wavefunction after evaluating all the loop diagrams, \figs\ref{fig:selfenergy}, \ref{fig:selfenergyinst}, \ref{fig:vertexT} and \ref{fig:instvertex}, can be decomposed in terms of four structures
\begin{multline}\label{eq:genvertex}
\bar u(0) \epsl_{\lambda}(q)v(1)\left [1 + \left (\frac{\alpha_s\cf}{2\pi}\right )\mathcal{V}^{T}\right ] 
\\
+ \frac{q^+}{2k^+_0k^+_1}(\Pt \cdot \epst_{\lambda}) \bar u(0)\gamma^+ v(1) \left (\frac{\alpha_s\cf}{2\pi}\right )\mathcal{N}^{T}
\\
 +  \frac{q^+}{2k^+_0k^+_1} \frac{(\Pt \cdot \epst_{\lambda})}{\Pt^2}\Pt^j m \bar u(0)\gamma^+\gamma^j v(1) \left (\frac{\alpha_s\cf}{2\pi}\right )\mathcal{S}^{T} 
 \\ - \frac{q^+}{2k^+_0k^+_1}m \epst_{\lambda}^{j} \bar u(0)\gamma^+ \gamma^j  v(1)\left (\frac{\alpha_s\cf}{2\pi}\right )\mathcal{M}^{T}
, 
\end{multline}
where $\alpha_s =g^2/4\pi$ is the QCD coupling constant,  $\cf = (\nc^2-1)/(2\nc)$ and $\nc$ is the number of colors. We obtain the 4 separate scalar functions $\mathcal{V}^{T}$, $\mathcal{N}^{T}$, $\mathcal{S}^{T}$ and $\mathcal{M}^{T}$ by evaluating the loop diagrams. For the longitudinal photon, one can perform a similar, but simpler  decomposition.  Comparing \nr{eq:melem} and \nr{eq:genvertex} we can see that the vertex mass is related to  $\mathcal{M}^{T}$. 

\paragraph{On shell renormalization scheme}
For mass renormalization in the on-shell scheme we must look at the wave function in a specific kinematical configuration that we refer to as the \emph{on-shell point}, corresponding to a timelike virtual photon with $q^-= (q^+/(2k_0^+k_1^+))(\Pt^2+m^2)$ (note that the physical region for DIS is spacelike, $q^-<0$).
From Lorentz-invariance we know that at the on-shell point the whole $\gamma^* q\bar{q}$ vertex function can be expressed in terms of two known scalar functions, the Dirac and Pauli form factors
\begin{equation}
\hspace{-0.1cm} F_D(q^2/m^2) \bar u(0)\gamma^{\mu}v(1) 
+ F_P(q^2/m^2) \frac{iq_{\nu}}{2m}  \bar u(0)\sigma^{\mu\nu}v(1).
\end{equation}
It is a straightforward exercise to express $F_D(q^2/m^2)$ and $F_P(q^2/m^2)$ in terms of $\mathcal{V}^{T}$, $\mathcal{N}^{T}$, $\mathcal{S}^{T}$ and $\mathcal{M}^{T}$.

One mass renormalization condition is given by the requirement that the self-energy diagrams in  \figs\ref{fig:selfenergy} and~\ref{fig:selfenergyinst} do not have a pole at the on-shell point, as discussed explicitly in Ref.~\cite{Beuf:2021qqa}.  For a Lorentz-invariant regularization including the self-induced inertia diagrams, no other conditions are needed and the four conditions for $\mathcal{V}^{T}$, $\mathcal{N}^{T}$, $\mathcal{S}^{T}$ and $\mathcal{M}^{T}$ at the on-shell point are  additional nontrivial checks of our result. On the other hand, with the regularization scheme of Refs.~\cite{Beuf:2016wdz,Beuf:2017bpd,Hanninen:2017ddy}, the condition on $\mathcal{M}^{T}$ becomes an additional vertex mass renormalization condition and one is left with three consistency checks for  $\mathcal{V}^{T}$, $\mathcal{N}^{T}$, $\mathcal{S}^{T}$. In both cases our result for the mass-renormalized wave function is the same.

\paragraph{From wave function to cross section}
To calculate the inclusive DIS cross section, one additionally needs to specify the interaction of the state with the target proton or nucleus.  In the CGC formalism~\cite{Gelis:2010nm} this is described by  an eikonal interaction with a nonperturbatively strong color field. The field is parametrized in terms of Wilson lines as functions of the transverse coordinate. Thus one must, after performing the mass renormalization, transform the LCWF's into mixed transverse coordinate-longitudinal momentum space. The interactions of the mixed space states with the target bring in Wilson line correlators that are the same as in the massless case. Also similarly to the massless case, there are cancellations of divergences (appearing as $1/\varepsilon$ poles) and scheme-dependent terms between the $q\bar{q}$ and $q\bar{q}g$
contributions. In order to obtain a manifestly finite expression for the cross section these must be subtracted from the $q\bar{q}g$ terms and added to the 
$q\bar{q}$ terms. We are performing this step within the same subtraction scheme as in Ref.~\cite{Hanninen:2017ddy}.

\section{Result and discussion}

For high energy QCD calculations one needs the wavefunction in mixed transverse coordinate-longitudinal momentum space.  Some of the spinor matrix elements in \eq\nr{eq:genvertex} depend on the relative $q\bar{q}$ transverse momentum $\Pt$. Thus, what is needed are the scalar functions $\mathcal{V}^{T}$, $\mathcal{N}^{T}$, $\mathcal{S}^{T}$ and $\mathcal{M}^{T}$ multiplied by specific powers of the transverse momentum and by the leading order energy denominator, Fourier-transformed  to coordinate space. We denote this multiplication and transformation by $\mathcal{F}$.  The NLO $\gamma^* \to q\bar{q}$ LCWF, the main result of this paper,  can be written as
\begin{widetext}
\begin{equation}
\label{eq:finalNLOmixspacereduced}
\begin{split}
\widetilde{\psi}^{\gamma^{\ast}_T\rightarrow q\bar{q}}_{\nlo}  & = -\frac{ee_f}{2\pi}  \left (\frac{\alpha_s\cf}{2\pi}\right )  \Biggl \{ \Biggl [\left (\frac{k_0^+-k_1^+}{q^+} \right )\delta^{ij} \bar{u}(0)\gamma^{+}v(1)  + \frac{1}{2}\bar{u}(0)\gamma^{+}[\gamma^i, \gamma^j]v(1)  \Biggr ] \mathcal{F} \biggl [\Pt^i \mathcal{V}^T \biggr ] + \bar u(0)\gamma^+ v(1) \mathcal{F} \biggl [\Pt^j \mathcal{N}^T \biggr ]\\
& +   m \bar u(0)\gamma^+\gamma^i v(1) \mathcal{F} \biggl [\left (\frac{\Pt^i \Pt^j}{\Pt^2} - \frac{\delta^{ij}}{2}\right ) \mathcal{S}^T \biggr ]
- m \bar u(0)\gamma^+\gamma^ {j}v(1)\mathcal{F} \biggl [\mathcal{V}^T + \mathcal{M}^T - \frac{\mathcal{S}^T}{2} \biggr ]
\Biggr \}\epst^{j}_{\lambda}.
\end{split}
\end{equation}
The Fourier transformed scalar function $\mathcal{V}^T$ reads 
\begin{equation}
\begin{split}
    &\mathcal{F}\biggl [\Pt^i \mathcal{V}^T \biggr ]   = \frac{i\xt_{01}^i}{ \vert\xt_{01}\vert} \left (\frac{\kaz}{2\pi\vert \xt_{01}\vert} \right )^{\frac{D}{2}-2}\Biggl \{\biggl [\frac{3}{2} + \log \left ( \frac{\alpha}{z} \right ) + \log \left (\frac{\alpha}{1-z}\right ) \biggr ] \biggl \{\frac{(4\pi)^{2-\frac{D}{2}}}{(2-\frac{D}{2})}\Gamma\left (3 - \frac{D}{2}\right ) + \log\left (\frac{\vert \xt_{01}\vert^2 \mu^2}{4}\right )\\
    & + 2\gamma_E\biggr \} + \frac{1}{2}\frac{(D_s-4)}{(D-4)} \Biggr \} \kaz K_{\frac{D}{2}-1}\left (\vert \xt_{01}\vert \kaz\right ) + \frac{i\xt_{01}^i}{\vert \xt_{01}\vert} \Biggl \{\biggl [\frac{5}{2} - \frac{\pi^2}{3} + \log^2\left (\frac{z}{1-z} \right ) + \Omega_{\mathcal{V}}^T + L\biggr ] \kaz K_{1}\left (\vert \xt_{01}\vert \kaz \right )  + I_{\mathcal{V}}^T\Biggr \},
\end{split}
\end{equation}
where we have defined $\kappa_z = \sqrt{z(1-z)Q^2 + m^2}$. Here  $\vert \xt_{01}\vert = \vert \xt_0 - \xt_1 \vert $ is the transverse size of the $q\bar{q}$ dipole and $\mu^2$ is the transverse dimensional regularization scale. The factor $(D_s-4)/(D-4)$ is the regularization scheme dependent part, from which the FDH scheme result is obtained as $D_s \to 4$ and the CDR one as $D_s \to D = 4-2\varepsilon$. The function $K_\nu$ is the modified Bessel function of the second kind and the functions $\Omega_{\mathcal{V}}^T$ and $I_{\mathcal{V}}^T$ are given by:
\begin{eqnarray}
\label{eq:omegaVT}
\Omega_{\mathcal{V}}^T &=&    \left (1 + \frac{1}{2z} \right )\biggl [\log(1-z)  + \gamma \log \left (\frac{1 + \gamma}{1 + \gamma - 2z} \right ) \biggr ] - \frac{1}{2z}\biggl [ \left (z + \frac{1}{2} \right )(1-\gamma) +  \frac{m^2}{Q^2}\biggr ]\log \left (\frac{\kaz^2}{m^2} \right ) + [z \leftrightarrow 1-z]
\\
\label{eq:IVT}
  I_{\mathcal{V}}^T & =& \int_0^1 \frac{\ud \xi}{\xi} \left (\frac{2\log(\xi)}{(1-\xi)} - \frac{(1+\xi)}{2} \right ) \Biggl \{\sqrt{\kaz^2 + \frac{\xi(1-z)}{(1-\xi)}m^2}K_{1}\left (\vert \xt_{01}\vert \sqrt{\kaz^2 + \frac{\xi(1-z)}{(1-\xi)}m^2}\right ) - [\xi \to 0]\Biggr \}
  \\ \nonumber
 &&
 -\int_0^1 \ud \xi \left (\frac{\log(\xi)}{(1-\xi)^2} + \frac{z}{(1-x)} + \frac{z}{2} \right ) \frac{(1-z)m^2}{\sqrt{\kaz^2 + \frac{\xi(1-z)}{(1-\xi)}m^2}}K_{1}\left (\vert \xt_{01}\vert \sqrt{\kaz^2 + \frac{\xi(1-z)}{(1-\xi)}m^2}\right )
 \\ \nonumber
&& - \int_{0}^{z} \frac{\ud \chi}{(1-\chi)}
\int_{0}^{\infty}\frac{\ud u }{u(u + 1)} \frac{ m^2}{\kac^2} \biggl [2\chi + \left (\frac{u}{1+u}\right )^2 \frac{1}{z}(z-\chi)(1-2\chi) \biggr ]
\\ \nonumber
&& \hspace{2cm} \times \Biggl \{\sqrt{\kaz^2 + u\frac{(1-z)}{(1-\xi)}\kac^2}K_{1}\left (\vert \xt_{01}\vert \sqrt{\kaz^2 + u\frac{(1-z)}{(1-\xi)}\kac^2}\right ) - [u \to 0] \Biggr \}
\\ \nonumber
&& -  \int_{0}^{z} \frac{\ud \chi}{(1-\chi)^2}
\int_{0}^{\infty}\frac{\ud u}{(u + 1)} \, \left (z - \chi\right )\biggl [1 - \frac{2u}{1+u}\left (z - \chi\right ) +  \left (\frac{u}{1+u} \right )^2 \frac{1}{z}\left (z - \chi\right )^2 \biggr ]
\\ \nonumber
&& \hspace{2cm}\times \frac{m^2}{\sqrt{\kaz^2 + u\frac{(1-z)}{(1-\chi)}\kac^2}}K_{1}\left (\vert \xt_{01}\vert \sqrt{\kaz^2 + u\frac{(1-z)}{(1-\chi)}\kac^2}\right )  \quad +  \quad \big[z \leftrightarrow 1-z\big].
\end{eqnarray}
Here $+[z \leftrightarrow 1-z]$ adds a term corresponding to  the whole preceding expression with the replacement.
The Fourier transformed scalar function $\mathcal{N}^T$ reads
\begin{equation}
    \mathcal{F} \biggl [\Pt^j \mathcal{N}^T \biggr ] = \frac{i\xt_{01}^j}{\vert \xt_{01}\vert}  \Biggl \{ \Omega_{\mathcal{N}}^T \,  \kaz  K_{1}\left (\vert \xt_{01}\vert \kaz \right ) + I_{\mathcal{N}}^T  \Biggr \},
\end{equation}
where the functions $\Omega_{\mathcal{N}}^T$ and $I_{\mathcal{N}}^T$ are given by:
\begin{eqnarray}
\label{eq:omegaNT}
\Omega_{\mathcal{N}}^T &=&  
\frac{z+1-2z^2}{z}
\Biggl [\log(1-z)  + \gamma \log \left (\frac{1+\gamma}{1+\gamma-2z}  \right ) \Biggr ] - \frac{(1-z)}{z} \Biggl [ 
\frac{2z+1}{2}
(1-\gamma ) + \frac{m^2}{Q^2} \Biggr ]\log \left (\frac{\kaz^2}{m^2} \right ) - [z \leftrightarrow 1-z]
\\
\label{eq:INT}
  I_{\mathcal{N}}^T   &=& \frac{2(1-z)}{z} \int_0^z \ud \chi \int_0^\infty \frac{\ud u}{(u+1)^3} \Biggl \{\left [(2+u)u z + u^2\chi\right ]\sqrt{\kaz^2 + u \frac{(1-z)}{(1-\chi)}\kac^2} \, K_1\left (\vert \xt_{01}\vert  \sqrt{\kaz^2 + u \frac{(1-z)}{(1-\chi)}\kac^2} \right ) 
  \\
&&
\nonumber
\!\!\!\!\!\!\!\! \!\!\!\!+ \frac{m^2}{\kac^2} 
\biggl (\frac{z}{1-z} + \frac{\chi}{1 - \chi} [u  - 2z - 2u\chi] \biggr )
\Biggl [\sqrt{\kaz^2 + u \frac{(1-z)}{(1-\chi)}\kac^2} \, K_1\left (\vert \xt_{01}\vert  \sqrt{\kaz^2 + u \frac{(1-z)}{(1-\chi)}\kac^2} \right ) -[u\to 0] \Biggr ]\Biggr \}
- [z \leftrightarrow 1-z].
\end{eqnarray}
The Fourier transformed scalar function $\mathcal{S}^T$ reads 
\begin{equation}
\begin{split}
    \mathcal{F} \biggl [\left (\frac{\Pt^i \Pt^j}{\Pt^2} - \frac{\delta^{ij}}{2}\right )\mathcal{S}^T \biggr ] & = \frac{(1 - z)}{2} \left[\frac{\xt_{01}^i\xt_{01}^j}{\vert \xt_{01}\vert^2}- \frac{\delta^{ij}}{2}\right]
\int_{0}^{z} \frac{\ud \chi}{(1 - \chi)}\,
\int_{0}^{\infty}\frac{\ud u}{(u + 1)^2} \,
\vert \xt_{01}\vert  \sqrt{\kaz^2 + u \frac{(1 - z)}{(1 - \chi)} \kac^2}
\\
& \hspace{1cm}
\times 
K_1\left(\vert \xt_{01}\vert \sqrt{\kaz^2 + u \frac{(1 - z)}{(1 - \chi)} \kac^2}\right) + [z \leftrightarrow 1-z].
\end{split}
\end{equation}
Finally, the Fourier transformed scalar function combination $\mathcal{V}^T + \mathcal{M}^T - \mathcal{S}^T/2$ reads
\begin{equation}
\begin{split}
& \mathcal{F}\biggl [\mathcal{V}^T + \mathcal{M}^T - \frac{\mathcal{S}^T}{2} \biggr ]  = \left (\frac{\kaz}{2\pi\vert \xt_{01}\vert} \right )^{\frac{D}{2}-2}\!\!\Biggl \{\biggl [\frac{3}{2} + \log\left ( \frac{\alpha}{z}\right ) +  \log\left ( \frac{\alpha}{1-z}\right ) \biggr ]\biggl \{\frac{(4\pi)^{2-\frac{D}{2}}}{(2-\frac{D}{2})}\Gamma \left (3 - \frac{D}{2}\right ) + \log \left (\frac{\vert \xt_{01}\vert^2 \mu^2}{4}\right )\\
& + 2\gamma_E \biggr \} +  \frac{1}{2}\frac{(D_s-4)}{(D-4)} \Biggr \}K_{\frac{D}{2}-2}\left (\vert \xt_{01}\vert \kaz \right ) +  \Biggl \{ 3 - \frac{\pi^2}{3}  + \log^2 \left (\frac{z}{1-z} \right )  + \Omega_{\mathcal{V}}^T + L  \Biggr \}K_{0}\left (\vert \xt_{01}\vert \kaz \right )   + I_{\mathcal{VMS}}^T, 
\end{split}    
\end{equation}
where the function $I_{\mathcal{VMS}}^T$ is given by
\begin{equation}
\label{eq:IVMST}
\begin{split}
I_{\mathcal{VMS}}^T  & = \int_{0}^{1}\frac{\ud \xi}{\xi} \left (\frac{2\log(\xi)}{(1-\xi)} - \frac{(1+\xi)}{2}\right )\Biggl \{ K_{0}\left (\vert \xt_{01}\vert \sqrt{\kaz^2  + \frac{\xi(1-z)}{(1-\xi)}m^2}\right )  -[\xi \to 0] \Biggr \}\\
& + \int_{0}^{1}\ud \xi \left (-\frac{3(1-z)}{2(1-\xi)} + \frac{(1-z)}{2}\right ) K_{0}\left (\vert \xt_{01}\vert \sqrt{\kaz^2  + \frac{\xi(1-z)}{(1-\xi)}m^2}\right )\\
& + \int_{0}^{z} \frac{\ud \chi}{(1-\chi)}\int_{0}^{\infty} \frac{\ud u}{(u+1)^2} \biggl \{ -z - \frac{u}{(1+u)}\frac{(z +u\chi)}{z}(\chi  - (1-z)) \biggr \} 
K_{0}\left (\vert \xt_{01}\vert \sqrt{\kaz^2  + u\frac{(1-z)}{(1-\chi)}\kac^2}\right )\\
& + \int_{0}^{z} \ud \chi \int_{0}^{\infty} \frac{\ud u}{(u+1)^3}  \biggl \{\frac{\kaz^2}{\kac^2} \biggl [1 + u\frac{\chi(1-\chi)}{z(1-z)}\biggr ] - \frac{m^2}{\kac^2} \frac{\chi}{(1-\chi)}\bigg [2\frac{(1+u)^2}{u} +   \frac{u}{z(1-z)}\left (z - \chi\right )^2 \biggr ]\biggr \}\\
&  \hspace{2cm}  \times \Biggl \{K_{0}\left (\vert \xt_{01}\vert \sqrt{\kaz^2  + u\frac{(1-z)}{(1-\chi)}\kac^2}\right ) - [u \to 0]    \Biggr \} + [z \leftrightarrow 1-z]. 
\end{split}
\end{equation}

\end{widetext}
The above expressions use the function  $L$, which is defined as 
\begin{equation}
\label{eq:Lfunction}
L = \sum_{\sigma= \pm 1} \biggl [ \mathrm{Li}_2 \left (\frac{1}{1-\frac{1}{2z}(1+\sigma \gamma)} \right ) + [z \leftrightarrow 1-z]
\biggr ],
\end{equation}
where $\mathrm{Li}_2$ is the standard dilogarithm function and the parameter $\gamma = \sqrt{1 + 4m^2/Q^2}$. To our knowledge this is a completely new fundamental result in perturbative QCD.

We have also calculated the total DIS cross section to one loop order, using our results for the $q\bar{q}$ LCWF and the more straightforward, but algebraically complicated gluon emission wave functions. After the cancellation of UV divergences between the $q\bar{q}$ and $q\bar{q}g$ contributions, 
the cross section has a similar structure as for massless quarks~\cite{Beuf:2017bpd,Hanninen:2017ddy}:
\begin{equation}
\label{eq:xs}
\sigma^{\gamma^{\ast}}_{L,T} = \sigma^{\gamma^{\ast}}_{L,T}\bigg\vert^{\rm subt}_{q\bar{q}} +  \sigma^{\gamma^{\ast}}_{L,T}\bigg\vert^{\rm subt}_{q\bar{q}g} + \mathcal{O}(\alpha_{em}\alpha_s^2),
\end{equation}
where $\alpha_{em}$ is the QED coupling constant. The ``dipole'' contribution $\left. \sigma^{\gamma^{\ast}}_{L,T}\right\vert^{\rm subt}_{q\bar{q}}$ corresponds to just the quark-antiquark pair crossing the shockwave color field of the target. The $q\bar{q}g$-term 
$\left. \sigma^{\gamma^{\ast}}_{L,T}\right\vert^{\rm subt}_{q\bar{q}g}$ corresponds to a quark-antiquark-gluon system crossing the shockwave. 
The integration in the limit $k^+ \to 0$ for this gluon develops a logarithmically large contribution, which must be resummed into the B/JIMWLK evolution of the target color fields in the same way as for massless quarks~\cite{Ducloue:2017ftk,Beuf:2020dxl}. The transverse coordinate and gluon momentum fraction integrals cannot be performed analytically in the general case, since they depend on the properties of the Wilson line correlators describing the target. The quark momentum fraction integrals are also best left for  numerical evaluation, similarly as in the case of massless quarks.

In addition to  integrals that are similar to the massless case, the mass-dependent parts include additional integrals over Schwinger parameters that we have not been able to perform analytically. These integrals are generalizations of Bessel $K_{0,1}$-function integral representations appearing in the massless case. They are very well convergent, and we do not expect their numerical evaluation to be significantly more complicated than a numerical evaluation of a normal Bessel $K_{0,1}$-function. All the explicit expressions of the cross sections are written out in the Supplemental material of this Letter.

In conclusion, after a lengthy calculation, we have obtained the one loop LCWF's for the process $\gamma^*\to q \bar{q}$. These are new results in field theory by themselves, expressing the full one-loop structure of the photon-quark-antiquark vertex in light cone gauge. We believe our result will be an important element in many future calculations. For example, the LCWF's will enable several calculations of exclusive processes in high energy DIS, such as diffractive structure functions, diffractive dijets, and exclusive vector meson production, at NLO accuracy and including massive quarks. 
As a first important application, we have computed the full NLO cross section for DIS in the dipole picture with quark masses. The cross section expressions obtained in this work will pave the way for simultaneous global fits of total and heavy quark cross sections measured at HERA, following the massless quark case~\cite{Beuf:2020dxl}. These cross sections will be crucial for obtaining more precise predictions for EIC cross sections including the effects of gluon saturation.

\section*{Acknowledgments} 

We thank H. M\"{a}ntysaari, J. Penttala and H. H\"{a}nninen for useful discussions. This work has been supported by the Academy of Finland, projects  321840 and 1322502, under the European Union’s Horizon 2020 research and innovation programme by the STRONG-2020 project (grant agreement No 824093), 
by the European Research Council,  grant agreements ERC-2015-CoG-681707 and ERC-2016-CoG-725369,
and by the National Science Centre (Poland) under the research Grant No. 2020/38/E/ST2/00122 (SONATA BIS 10).  The content of this article does not reflect the official opinion of the European Union and responsibility for the information and views expressed therein lies entirely with the authors.

\appendix

\section{Supplemental material}

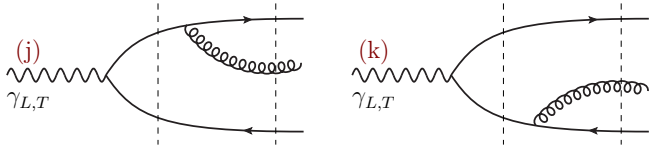
\begin{figure}[tbh]
\labeldiag{diag:qgqbar}
\labeldiag{diag:qqbarg}
\centerline{
\resizebox{0.25\textwidth}{!}{
\includegraphics[width=5.4cm]{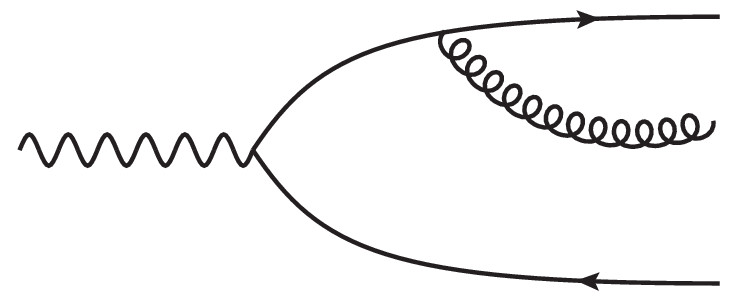}
\begin{tikzpicture}[overlay]
\draw [dashed] (-2.8,2.3) -- (-2.8,-0.1);
\draw [dashed] (-0.8,2.3) -- (-0.8,-0.1);
\node[anchor=north] at (-5,0.9) {\large $\gamma_{L,T}$};
\node[anchor=north] at (-5,1.8) {\large\ref{diag:qgqbar}};
 \end{tikzpicture}
}
\resizebox{0.25\textwidth}{!}{
\includegraphics[width=5.4cm]{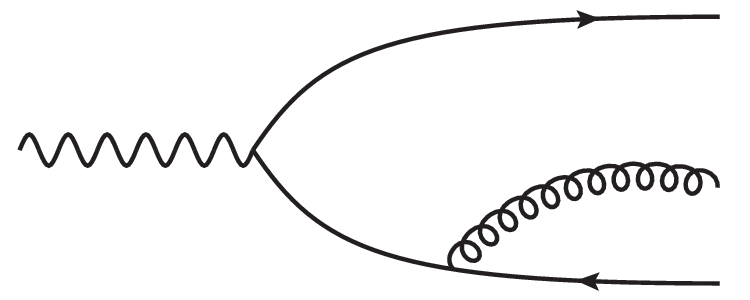}
\begin{tikzpicture}[overlay]
\draw [dashed] (-2.8,2.3) -- (-2.8,-0.1);
\draw [dashed] (-0.8,2.3) -- (-0.8,-0.1);
\node[anchor=north] at (-5,0.9) {\large $\gamma_{L,T}$};
\node[anchor=north] at (-5,1.8) {\large\ref{diag:qqbarg}};
 \end{tikzpicture}
}
}
\caption{Gluon emission diagrams contributing to the $\gamma^*\to q\bar{q}g$ LCWF.}
\label{fig:qgqbar}
 \end{figure}

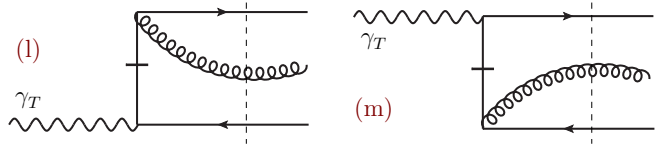
\begin{figure}[tbh]
\labeldiag{diag:qgqbarinst}
\labeldiag{diag:qqbarginst}
\centerline{
\resizebox{0.25\textwidth}{!}{
\includegraphics[width=5.4cm]{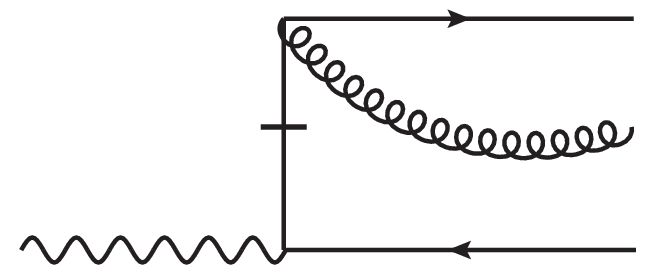}
\begin{tikzpicture}[overlay]
\draw [dashed] (-1.3,2.3) -- (-1.3,-0.1);
\node[anchor=south] at (-5,0.4) {\large $\gamma_{T}$};
\node[anchor=north] at (-5,1.8) {\large\ref{diag:qgqbarinst}};
 \end{tikzpicture}
}
\resizebox{0.25\textwidth}{!}{
\includegraphics[width=5.4cm]{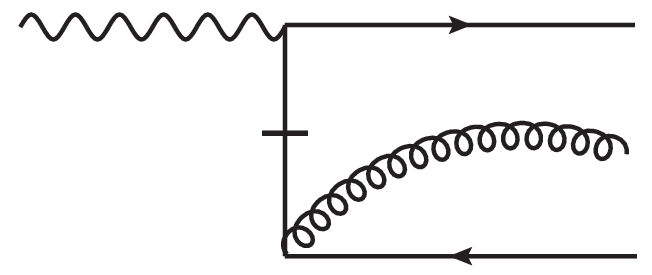}
\begin{tikzpicture}[overlay]
\draw [dashed] (-1.3,2.3) -- (-1.3,-0.1);
\node[anchor=north] at (-5,1.9) {\large  $\gamma_{T}$};
\node[anchor=north] at (-5,0.8) {\large\ref{diag:qqbarginst}};
 \end{tikzpicture}
}
}
\caption{Gluon emission diagrams with an instantaneous vertex, only contributing to the transverse photon wave function.}
\label{fig:qgqbarinst}
 \end{figure}

Here we present the results for the transverse and longitudinal photon cross sections, where the latter was already published in \cite{Beuf:2021qqa} but is here shown in the notations of this paper. These require also the contribution of the radiative diagrams shown in Figs.~\ref{fig:qgqbar} and~\ref{fig:qgqbarinst}.

\begin{widetext}

\section{Explicit expressions for transverse photon cross section}

Using the generic notation \nr{eq:xs} the (subtracted) quark-antiquark ($q\bar{q}$) contribution to the transverse photon total DIS cross section at NLO in $\as$ with massive quarks reads
\begin{equation}
\begin{split}
 \sigma^{\gamma^{\ast}}_{T}\bigg\vert^{\rm subt}_{q\bar{q}} & =   4\nc\alpha_{em}\sum_{f}e_f^2\int_{0}^{1} \ud z \, \int_{\xt_0}\int_{\xt_1} \,  \Biggl \{ \biggl [z^2 + (1-z)^2 \biggr ]  [\kaz K_1(\vert \xt_{01}\vert \kaz)]^2  + m^2 [K_0(\vert \xt_{01}\vert \kaz)]^2\\
& + \left(\frac{\alpha_s\cf}{\pi}\right ) \biggl \{\biggl [z^2 + (1-z)^2 \biggr ] \kaz K_1(\vert \xt_{01}\vert \kaz)f_{\mathcal{V}}^T + \frac{(2z-1)}{2} \kaz K_1(\vert \xt_{01}\vert \kaz)f_{\mathcal{N}}^T\\
& \hspace{2cm} + m^2 K_0(\vert \xt_{01}\vert \kaz) f_{\mathcal{VMS}}^T \biggr \} \Biggr \}\mathrm{Re}[1-\mathcal{S}_{01}],
\end{split}
\end{equation}
where $\mathcal{S}_{01}$ is the quark-antiquark dipole amplitude (see the definition in Ref.~\cite{Beuf:2021qqa}),  $Q$ is the virtuality of the photon, $z = k_0^+/q^+$  is longitudinal momentum fraction, $m$ is the quark mass and the short-hand notation  $\int_{\xt_i} = \int \ud^2 \xt_i/(2\pi)$ for $i=0,1$ denotes a 2-dimensional transverse coordinate integral. The functions $f_{i}^T$ which encode the one loop QCD corrections with massive quarks are given by the following expressions:
\begin{equation}
\begin{split}
f_{\mathcal{V}}^T & = \biggl \{\frac{5}{2} - \frac{\pi^2}{3} + \log^2\left (\frac{z}{1-z}\right ) + \Omega_{\mathcal{V}}^T + L  \biggr \} \kaz K_1(\vert \xt_{01}\vert \kaz) + \mathcal{I}_{\mathcal{V}}^T, \\
f_{\mathcal{N}}^T & = \Omega_{\mathcal{N}}^T\, \kaz K_1(\vert \xt_{01}\vert \kaz) + \mathcal{I}_{\mathcal{N}}^T, \\
f_{\mathcal{VMS}}^T & = \biggl \{3 - \frac{\pi^2}{3} + \log^2\left (\frac{z}{1-z}\right )  + \Omega_{\mathcal{V}}^T + L \biggr \}K_0(\vert \xt_{01}\vert \kaz) +  \mathcal{I}_{\mathcal{VMS}}^T.
\end{split}    
\end{equation}
Here the function $L$ is defined in \eq\nr{eq:Lfunction} and the functions $\Omega_{\mathcal{V}}^T, \mathcal{I}_{\mathcal{V}}^T, \Omega_{\mathcal{N}}^T, \mathcal{I}_{\mathcal{N}}^T$ and $\mathcal{I}_{\mathcal{VMS}}^T$ are given by \eqs\nr{eq:omegaVT}, \nr{eq:IVT}, \nr{eq:omegaNT}, \nr{eq:INT} and \nr{eq:IVMST}, respectively.

The (subtracted) quark-antiquark-gluon ($q\bar{q}g$) contribution to the transverse photon total DIS cross section
at NLO in $\as$ with massive quarks can be written as a sum of four contribution
\begin{equation}
\label{eq:qbarqgtotDIS}
\sigma^{\gamma^{\ast}}_{T}\bigg\vert^{\rm subt}_{q\bar qg} =   \sigma^{\gamma^{\ast}}_{T}\bigg\vert^{ {\rm UV-subt}}_{q\bar qg} + \sigma^{\gamma^{\ast}}_{T}\bigg\vert^{ {\rm UV_{m}-subt}}_{q\bar qg} +  \sigma^{\gamma^{\ast}}_{T}\bigg\vert^{ {\rm F}}_{q\bar qg}  +  \sigma^{\gamma^{\ast}}_{T}\bigg\vert^{ {\rm F}_m}_{q\bar qg},
\end{equation}
where the first and the second term in \eq\nr{eq:qbarqgtotDIS} are the UV subtracted parts and the last two terms are the UV finite contribution. 

It is useful in the following to introduce the compact notation for coordinate differences
\begin{equation}
\label{eq:xtpnmdef}
\xt_{n+m;p} = -\xt_{p;n+m} = \left ( \frac{k^+_n\xt_n + k^+_m\xt_m}{k^+_n + k^+_m} \right ) - \xt_p
\end{equation}
with $\xt_{nm} = \xt_n - \xt_m$. 
It is also useful to introduce the relative coordinates corresponding to diagrams \ref{diag:qgqbar} and \ref{diag:qqbarg} as $\xt_{3;\ref{diag:qgqbar}}=  \xt_{0+2;1}$, $\xt_{3;\ref{diag:qqbarg}}=  \xt_{0;1+2}$,
$\xt_{2;\ref{diag:qgqbar}}=  \xt_{20}$ and $\xt_{2;\ref{diag:qqbarg}}=  \xt_{21}$. Using this notation, the UV subtracted contributions can be written as
\begin{equation}
\label{eq:exptsubtqbarqghnf}
\begin{split}
&\sigma^{\gamma^{\ast}}_{T}\bigg\vert^{\rm UV-subt}_{q\bar qg}   =  4\nc\alpha_{em} \left (\frac{\alpha_s\cf}{\pi}\right )\sum_{f}e_f^2\int_{\xt_0}\int_{\xt_1}\int_{\xt_2} \int_{0}^{\infty} \ud k^+_0\int_{0}^{\infty} \ud k^+_1\int_{0}^{\infty} \frac{\ud k^+_2}{k^+_2}\frac{\delta(q^+-\sum_{i=0}^{2}k^+_i)}{q^+}\\
&\times \Biggl \{\frac{1}{(k_0^+ + k_2^+)^2}\biggl [2k^+_0(k^+_0 + k^+_2) + (k^+_2)^2 \biggr ]\biggl [1 - \frac{2k_1^+(q^+-k_1^+)}{(q^+)^2} \biggr ]\\
&  \times \biggl \{   \frac{\vert \xt_{3;\ref{diag:qgqbar}}\vert^2\vert\xt_{2;\ref{diag:qgqbar}}\vert^2}{256}[\mathcal{G}_{\ref{diag:qgqbar}}^{(2;2)}]^2 \,\mathrm{Re}[1-\mathcal{S}_{012}]  - \frac{e^{-\vert\xt_{2;\ref{diag:qgqbar}}\vert^2/(\vert\xt_{01}\vert^2 e^{\gamma_E})    }}{\vert\xt_{2;\ref{diag:qgqbar}}\vert^2} \biggl [\sqrt{\overline{Q}^2_{\ref{diag:qgqbar}} + m^2}K_1 \left (\vert \xt_{01}\vert \sqrt{\overline{Q}^2_{\ref{diag:qgqbar}} + m^2} \right )\biggr ]^2   \mathrm{Re}[1-\mathcal{S}_{01}]           \biggr \} \\
& + \frac{1}{(k_1^+ + k_2^+)^2}\biggl [2k^+_1(k^+_1 + k^+_2) + (k^+_2)^2 \biggr ]\biggl [1 - \frac{2k_0^+(q^+-k_0^+)}{(q^+)^2} \biggr ]\\
&  \times \biggl \{    \frac{\vert \xt_{3;\ref{diag:qqbarg}}\vert^2\vert\xt_{2;\ref{diag:qqbarg}}\vert^2}{256}[\mathcal{G}_{\ref{diag:qqbarg}}^{(2;2)}]^2 \,\mathrm{Re}[1-\mathcal{S}_{012}]  - \frac{e^{-\vert\xt_{2;\ref{diag:qqbarg}}\vert^2/(\vert\xt_{01}\vert^2 e^{\gamma_E})  }}{\vert\xt_{2;\ref{diag:qqbarg}}\vert^2} \biggl [\sqrt{\overline{Q}^2_{\ref{diag:qqbarg}} + m^2}  K_1 \left (\vert \xt_{01}\vert \sqrt{\overline{Q}^2_{\ref{diag:qqbarg}} + m^2} \right )\biggr ]^2   \mathrm{Re}[1-\mathcal{S}_{01}]   \biggr \}\Biggr \}
\end{split}
\end{equation}
and
\begin{equation}
\label{eq:exptsubtqbarqghf}
\begin{split}
&\sigma^{\gamma^{\ast}}_{T}\bigg\vert^{{\rm UV_{m}-subt}}_{q\bar qg}   =  4m^2\,\nc\alpha_{em} \left (\frac{\alpha_s\cf}{\pi}\right )\sum_{f}e_f^2\int_{\xt_0}\int_{\xt_1}\int_{\xt_2} \int_{0}^{\infty} \ud k^+_0\int_{0}^{\infty} \ud k^+_1\int_{0}^{\infty} \frac{\ud k^+_2}{k^+_2}\frac{\delta(q^+-\sum_{i=0}^{2}k^+_i)}{q^+}\\
&\times \Biggl \{\frac{1}{(k_0^+ + k_2^+)^2}\biggl [2k^+_0(k^+_0 + k^+_2) + (k^+_2)^2 \biggr ]\\
&  \times \biggl \{   \frac{\vert\xt_{2;\ref{diag:qgqbar}}\vert^2}{64}[\mathcal{G}_{\ref{diag:qgqbar}}^{(1;2)}]^2 \,\mathrm{Re}[1-\mathcal{S}_{012}]  - \frac{e^{-\vert\xt_{2;\ref{diag:qgqbar}}\vert^2/(\vert\xt_{01}\vert^2 e^{\gamma_E})    }}{\vert\xt_{2;\ref{diag:qgqbar}}\vert^2} \biggl [K_0\left (\vert \xt_{01}\vert \sqrt{\overline{Q}^2_{\ref{diag:qgqbar}} + m^2} \right )\biggr ]^2   \mathrm{Re}[1-\mathcal{S}_{01}]           \biggr \} \\
& + \frac{1}{(k_1^+ + k_2^+)^2}\biggl [2k^+_1(k^+_1 + k^+_2) + (k^+_2)^2 \biggr ]\\
&  \times \biggl \{    \frac{\vert\xt_{2;\ref{diag:qqbarg}}\vert^2}{64}[\mathcal{G}_{\ref{diag:qqbarg}}^{(1;2)}]^2 \,\mathrm{Re}[1-\mathcal{S}_{012}]  - \frac{e^{-\vert\xt_{2;\ref{diag:qqbarg}}\vert^2/(\vert\xt_{01}\vert^2 e^{\gamma_E})  }}{\vert\xt_{2;\ref{diag:qqbarg}}\vert^2} \biggl [K_0 \left (\vert \xt_{01}\vert \sqrt{\overline{Q}^2_{\ref{diag:qqbarg}} + m^2} \right )\biggr ]^2   \mathrm{Re}[1-\mathcal{S}_{01}] \biggr \}\Biggr \},
\end{split}
\end{equation}
where $\mathcal{S}_{012}$ is the quark-antiquark-gluon amplitude (see the definition in Ref.~\cite{Beuf:2021qqa}). We have also introduced the generalized Bessel function integral
\begin{equation}
\label{eq:defGx}
\mathcal{G}_{\textnormal{(x)}}^{(a;b)} = \int_{0}^{\infty} \frac{\ud u}{u^a} \exp{\left (\!-u\bigg[\overline{Q}^2_{\textnormal{(x)}} + m^2\bigg]\right )} \exp{\left (-\frac{\vert\xt_{3;\textnormal{(x)}}\vert^2}{4u}\right )}\int_{0}^{u/\omega_{\textnormal{(x)}}} \frac{\ud t}{t^b} \exp{\left (\!-t\bigg[\omega_{\textnormal{(x)}}\lambda_{(x)}m^2\bigg]\right )}\exp{\left (-\frac{\vert\xt_{2;\textnormal{(x)}}\vert^2}{4t}\right )}
\end{equation}
with (x) being either \ref{diag:qgqbar} or \ref{diag:qqbarg}
and 
\begin{equation}
\label{eq:coeffqbarqg}
\begin{split}
\overline{Q}^2_{\ref{diag:qgqbar}} & = \frac{k^+_1(q^+-k^+_1)}{(q^+)^2}Q^2,\quad \lambda_{\ref{diag:qgqbar}} = \frac{k^+_1k^+_2}{q^+k^+_0}, \quad \omega_{\ref{diag:qgqbar}}  = \frac{q^+k^+_0k^+_2}{k^+_1(k^+_0 + k^+_2)^2}
\end{split}
\end{equation}
and
\begin{equation}
\label{eq:coeffqbarqg}
\begin{split}
\overline{Q}^2_{\ref{diag:qqbarg}} & = \frac{k^+_0(q^+-k^+_0)}{(q^+)^2}Q^2,\quad \lambda_{\ref{diag:qqbarg}} = \frac{k^+_0k^+_2}{q^+k^+_1}, \quad \omega_{\ref{diag:qqbarg}}  = \frac{q^+k^+_1k^+_2}{k^+_0(k^+_1 + k^+_2)^2}.
\end{split}
\end{equation}
We note that the integral \nr{eq:defGx}  could be seen as a generalization of the integral representation of Bessel functions that appear in the massless case \cite{Beuf:2017bpd,  Hanninen:2017ddy}. These integrals are very rapidly converging at both small and large values of the integration variable, and hence they should be well suited for numerical evaluation as is.

The UV finite contributions can be written into the following form 
 \begin{equation}
\begin{split}
\sigma&^{\gamma^{\ast}}_{T}\bigg\vert^{\rm F}_{q\bar{q}g}    =  4\nc\alpha_{em} \frac{1}{2}\left (\frac{\alpha_s\cf}{\pi}\right )\sum_{f}e_f^2\int_{\xt_0}\int_{\xt_1}\int_{\xt_2} \int_{0}^{\infty} \ud k^+_0\int_{0}^{\infty} \ud k^+_1\int_{0}^{\infty} \frac{\ud k^+_2}{k^+_2}\frac{\delta(q^+-\sum_{i=0}^{2}k^+_i)}{q^+} \\
&\times \Biggl \{\frac{1}{(k_0^+ +k_2^+)(k_1^+ + k_2^+)(q^+)^2}\biggl \{q^+ k_2^+(k_0^+ - k_1^+)^2\biggl [(\xt_{3;\ref{diag:qgqbar}} \cdot \xt_{2;\ref{diag:qgqbar}})(\xt_{3;\ref{diag:qqbarg}}\cdot \xt_{2;\ref{diag:qqbarg}}) - (\xt_{3;\ref{diag:qgqbar}}\cdot \xt_{2;\ref{diag:qqbarg}}  )(\xt_{3;\ref{diag:qqbarg}} \cdot \xt_{2;\ref{diag:qgqbar}}) \biggr ]\\
& - \biggl [k_1^+(k_0^+ + k_2^+) + k_0^+(k_1^+ + k_2^+)\biggr ]\biggl [k_0^+(k_0^+ + k_2^+) + k_1^+(k_1^+ + k_2^+)\biggr ] (\xt_{3;\ref{diag:qgqbar}}\cdot \xt_{3;\ref{diag:qqbarg}})(\xt_{2;\ref{diag:qgqbar}}\cdot \xt_{2;\ref{diag:qqbarg}})\biggr \}\frac{[\mathcal{G}_{\ref{diag:qgqbar}}^{(2;2)}][\mathcal{G}_{\ref{diag:qqbarg}}^{(2;2)}]}{64}\\
& -  \frac{(k_0^+ + k_2^+)k_1^+ k_2^+}{16(k_1^+ + k_2^+)^2 q^+}(\xt_{3;\ref{diag:qgqbar}}\cdot \xt_{2;\ref{diag:qgqbar}}) [\mathcal{G}_{\ref{diag:qgqbar}}^{(2;2)}]\mathcal{H}_{\ref{diag:qqbarg}}
+ \frac{(k_1^+ + k_2^+)k_0^+ k_2^+}{16(k_0^+ + k_2^+)^2 q^+}(\xt_{3;\ref{diag:qqbarg}}\cdot \xt_{2;\ref{diag:qqbarg}}) [\mathcal{G}_{\ref{diag:qqbarg}}^{(2;2)}]\mathcal{H}_{\ref{diag:qgqbar}}  \\
&   - \frac{(k_0^+)^2 k_1^+ k_2^+}{16(k_0 + k_2^+)^3q^+} (\xt_{3;\ref{diag:qgqbar}}\cdot \xt_{2;\ref{diag:qgqbar}})
[\mathcal{G}_{\ref{diag:qgqbar}}^{(2;2)}]\mathcal{H}_{\ref{diag:qgqbar}}
+ \frac{(k_1^+)^2 k_0^+ k_2^+}{16(k_1 + k_2^+)^3q^+} (\xt_{3;\ref{diag:qqbarg}}\cdot \xt_{2;\ref{diag:qqbarg}}) [\mathcal{G}_{\ref{diag:qqbarg}}^{(2;2)}]\mathcal{H}_{\ref{diag:qqbarg}}
\\
& + \frac{(k_0^+ k_2^+)^2}{8(k_0^+ + k_2^+)^4} 
[\mathcal{H}_{\ref{diag:qgqbar}}]^2 
+ \frac{(k_1^+ k_2^+)^2}{8(k_1^+ + k_2^+)^4}
[\mathcal{H}_{\ref{diag:qqbarg}}]^2
\Biggr \}\mathrm{Re}[1-\mathcal{S}_{012}]
\end{split}
\end{equation}
and
\begin{equation}
\begin{split}
&\sigma^{\gamma^{\ast}}_{T}\bigg\vert^{{\rm F_m}}_{q\bar{q}g}    =  4\nc\alpha_{em}  \frac{m^2}{2}\left (\frac{\alpha_s\cf}{\pi}\right )\sum_{f}e_f^2\int_{\xt_0}\int_{\xt_1}\int_{\xt_2} \int_{0}^{\infty} \ud k^+_0\int_{0}^{\infty} \ud k^+_1\int_{0}^{\infty} \frac{\ud k^+_2}{k^+_2}\frac{\delta(q^+-\sum_{i=0}^{2}k^+_i)}{q^+} \\
&\times \Biggl \{\frac{(k_2^+)^4}{64(k_0^+ + k_2^+)^4 (q^+)^2}\biggl [4k_1^+(k_1^+ - q^+) + 2(q^+)^2\biggr ] \vert\xt_{3;\ref{diag:qgqbar}}\vert^2 [\mathcal{G}_{\ref{diag:qgqbar}}^{(2;1)}]^2 - \frac{k_0^+k_1^+(k_2^+)^2}{16(k_0^+ + k_2^+)^3 q^+}  (\xt_{3;\ref{diag:qgqbar}}\cdot \xt_{2;\ref{diag:qgqbar}}) [\mathcal{G}_{\ref{diag:qgqbar}}^{(1;2)}][\mathcal{G}_{\ref{diag:qgqbar}}^{(2;1)}] \\
& + \frac{(k_2^+)^4}{64(k_1^+ + k_2^+)^4 (q^+)^2}\biggl [4k_0^+(k_0^+ - q^+) + 2(q^+)^2 \biggr ] \vert\xt_{3;\ref{diag:qqbarg}}\vert^2 [\mathcal{G}_{\ref{diag:qqbarg}}^{(2;1)}]^2  + \frac{k_0^+k_1^+(k_2^+)^2}{16(k_1^+ + k_2^+)^3 q^+}   (\xt_{3;\ref{diag:qqbarg}}\cdot \xt_{2;\ref{diag:qqbarg}}) [\mathcal{G}_{\ref{diag:qqbarg}}^{(1;2)}][\mathcal{G}_{\ref{diag:qqbarg}}^{(2;1)}] \\
& -\frac{1}{32(k_0^+ + k_2^+)(k_1^+ + k_2^+)}\biggl [(2k_0^+ + k_2^+)(2k_1^+ + k_2^+) + (k_2^+)^2\biggr ] (\xt_{2;\ref{diag:qgqbar}} \cdot \xt_{2;\ref{diag:qqbarg}}) [\mathcal{G}_{\ref{diag:qgqbar}}^{(1;2)}][\mathcal{G}_{\ref{diag:qqbarg}}^{(1;2)}]\\
& + \frac{ (k_2^+)^4}{32(k_0^+ + k_2^+)^2(k_1^+ + k_2^+)^2 (q^+)^2}\biggl [(2k_0^+ + k_2^+)(2k_1^+ + k_2^+) + (k_2^+)^2\biggr ] (\xt_{3;\ref{diag:qgqbar}} \cdot \xt_{3;\ref{diag:qqbarg}}) \mathcal{G}_{\ref{diag:qgqbar}}^{(2;1)}][\mathcal{G}_{\ref{diag:qqbarg}}^{(2;1)}]\\
& + \frac{m^2}{16} \frac{2(k_2^+)^4}{(k_0^+ + k_2^+)^4}[\mathcal{G}_{\ref{diag:qgqbar}}^{(1;1)}]^2  + \frac{m^2}{16} \frac{2(k_2^+)^4}{(k_1^+ + k_2^+)^4} [\mathcal{G}_{\ref{diag:qqbarg}}^{(1;1)}]^2\\
& - \frac{(k_0^+k_2^+)^2}{16(k_0^+ + k_2^+)(k_1^+ + k_2^+)^2 q^+}  (\xt_{2;\ref{diag:qgqbar}} \cdot \xt_{3;\ref{diag:qqbarg}}) [\mathcal{G}_{\ref{diag:qgqbar}}^{(1;2)}][\mathcal{G}_{\ref{diag:qqbarg}}^{(2;1)}] + \frac{(k_1^+k_2^+)^2}{16(k_1^+ + k_2^+)(k_0^+ + k_2^+)^2 q^+}  (\xt_{3;\ref{diag:qgqbar}} \cdot \xt_{2;\ref{diag:qqbarg}}) [\mathcal{G}_{\ref{diag:qgqbar}}^{(2;1)}][\mathcal{G}_{\ref{diag:qqbarg}}^{(1;2)}]\\
& - \frac{(k_0^+k_1^+)(k_2^+)^2}{16(k_0^+ + k_2^+)^3 q^+} (\xt_{2;\ref{diag:qgqbar}} \cdot \xt_{3;\ref{diag:qgqbar}}) [\mathcal{G}_{\ref{diag:qgqbar}}^{(2;2)}][\mathcal{G}_{\ref{diag:qgqbar}}^{(1;1)}] + \frac{(k_0^+k_1^+)(k_2^+)^2}{16(k_1^+ + k_2^+)^3 q^+} (\xt_{2;\ref{diag:qqbarg}} \cdot \xt_{3;\ref{diag:qqbarg}}) [\mathcal{G}_{\ref{diag:qqbarg}}^{(2;2)}][\mathcal{G}_{\ref{diag:qqbarg}}^{(1;1)}]\\
& - \frac{(k_0^+ + k_2^+)(k_2^+)^2}{16(k_1^+ + k_2^+)^2 q^+} (\xt_{2;\ref{diag:qgqbar}} \cdot \xt_{3;\ref{diag:qgqbar}}) [\mathcal{G}_{\ref{diag:qgqbar}}^{(2;2)}][\mathcal{G}_{\ref{diag:qqbarg}}^{(1;1)}] 
+ \frac{(k_1^+ + k_2^+)(k_2^+)^2}{16(k_0^+ + k_2^+)^2 q^+}  (\xt_{2;\ref{diag:qqbarg}} \cdot \xt_{3;\ref{diag:qqbarg}}) [\mathcal{G}_{\ref{diag:qqbarg}}^{(2;2)}][\mathcal{G}_{\ref{diag:qgqbar}}^{(1;1)}]
\\
& + \frac{k_0^+(k_2^+)^3}{4(k_0^+ + k_2^+)^4}  \mathcal{H}_{\ref{diag:qgqbar}}[\mathcal{G}_{\ref{diag:qgqbar}}^{(1;1)}] + \frac{k_1^+(k_2^+)^3}{4(k_1^+ + k_2^+)^4}  \mathcal{H}_{\ref{diag:qqbarg}}[\mathcal{G}_{\ref{diag:qqbarg}}^{(1;1)}]\Biggr \}\mathrm{Re}[1-\mathcal{S}_{012}].
\end{split}
\end{equation}
Here we have also introduced the function 
$\mathcal{H}_{\textnormal{(x)}}$ defined as
\begin{equation}
\mathcal{H}_{\textnormal{(x)}} = \int_{0}^{\infty} \frac{\ud u}{u^2} \exp{\left (-u  \bigg [\overline{Q}^2_{\textnormal{(x)}} + m^2 + \lambda_{\textnormal{(x)}}m^2 \biggr ]\right )}\exp{\left (-\frac{
\vert \xt_{3;\textnormal{(x)}}\vert^2 + \omega_{\textnormal{(x)}}\vert\xt_{2;\textnormal{(x)}}\vert^2}{4u }\right )}
\end{equation}
with (x) being \ref{diag:qgqbar} or \ref{diag:qqbarg}.

\section{Explicit expressions for longitudinal photon cross section}

Using the generic notation \nr{eq:xs} the (subtracted) quark-antiquark ($q\bar{q}$) contribution to the longitudinal photon total DIS cross section at NLO in $\as$ with massive quarks reads \cite{Beuf:2021qqa}
\begin{equation}
\begin{split}
 \sigma^{\gamma^{\ast}}_{L}\bigg\vert^{\rm subt}_{q\bar{q}} \!\! =  4\nc\alpha_{em}4Q^2\sum_{f}e_f^2\int_{0}^{1} \ud z  [z(1-z)]^2\int_{\xt_0}\int_{\xt_1} \Biggl \{[K_0(\vert \xt_{01}\vert \kaz)]^2  + \left (\frac{\alpha_s\cf}{\pi}\right ) K_0(\vert \xt_{01}\vert \kaz)\, f_{\mathcal{V}}^L \Biggr \}\mathrm{Re}[1-\mathcal{S}_{01}],
\end{split}
\end{equation}
where the function $f_{\mathcal{V}}^L$ can be written as 
\begin{equation}
f_{\mathcal{V}}^L = \biggl \{ \frac{5}{2} - \frac{\pi^2}{3} + \log^2\left (\frac{z}{1-z}\right )   + \Omega_{\mathcal{V}}^L + L \biggr \}K_0(\vert \xt_{01}\vert \kaz) + I_{\mathcal{V}}^L.
\end{equation}
The functions $\Omega_{\mathcal{V}}^L$ and $\mathcal{I}_{\mathcal{V}}^L$ are given by the following expressions:
\begin{equation}
\label{eq:omegavL}
\begin{split}
\Omega_{\mathcal{V}}^L  =   \frac{1}{2z}\biggl [\log(1-z) &+ \gamma \log\left (\frac{1+\gamma}{1+\gamma - 2z}\right ) \biggr ]  + \frac{1}{2(1-z)}\biggl [\log(z) + \gamma \log\left (\frac{1+\gamma}{1+\gamma - 2(1-z)}\right ) \biggr ] \\
& + \frac{1}{4z(1-z)}\biggl[\left ( \gamma - 1\right ) + \frac{2m^2}{Q^2}\biggr ]\log\left (\frac{\overline{Q}^2 + m^2}{m^2}\right )
\end{split}
\end{equation}
and 
\begin{equation}
\label{eq:IVL}
\begin{split}
I_{\mathcal{V}}^L & =  -\int_{0}^{1} \frac{\ud \xi}{\xi}\biggl (-\frac{2\log(\xi)}{(1-\xi)} + \frac{(1+\xi)}{2}\biggr )\Biggl \{K_0\left (\vert \xt_{01}\vert\sqrt{\kaz^2 + \xi\frac{(1-z)}{1-\xi}m^2}\right ) - [\xi \to 0]\Biggr \}\\
& -\int_{0}^{z}\frac{\ud \chi}{(1-\chi)}
 \int_{0}^{\infty}\frac{\ud u}{(u+1)^2} \,\frac{m^2}{\kac^2}
\left[
\frac{2\chi (u+1)}{u} +\frac{(2z-1)\chi(z-\chi)}{z(1-z)}
-\frac{u(z-\chi)^2}{z(1-z)(u+1)}
\right]
\\
&  \hspace{4.5cm}\times
 \left\{K_0\left (\vert \xt_{01}\vert\sqrt{
\kaz^2 
+u\frac{(1-z)}{(1-\chi)}\kac^2 
}\right ) - [u \to 0] \right\} + [z \leftrightarrow 1-z].
\end{split}
\end{equation}

The (subtracted) quark-antiquark-gluon ($q\bar{q}g$) contribution to the longitudinal photon total DIS cross section
at NLO with massive quarks can be written as \cite{Beuf:2021qqa}
\begin{equation}
\begin{split}
&\sigma^{\gamma^{\ast}}_{L}\bigg\vert^{\rm subt}_{q\bar{q}g}    =  4\nc\alpha_{em} 4Q^2\left (\frac{\alpha_s\cf}{\pi}\right )\sum_{f}e_f^2\int_{\xt_0}\int_{\xt_1}\int_{\xt_2} \int_{0}^{\infty} \ud k^+_0\int_{0}^{\infty} \ud k^+_1\int_{0}^{\infty} \frac{\ud k^+_2}{k^+_2}\frac{\delta(q^+-\sum_{i=0}^{2}k^+_i)}{(q^+)^5} \\
&\times \Biggl \{(k^+_1)^2\biggl [2k^+_0(k^+_0 + k^+_2) + (k^+_2)^2 \biggr ]\\
& \hspace{1.5cm} \times \biggl \{   \frac{\vert\xt_{2;\ref{diag:qgqbar}}\vert^2}{64}[\mathcal{G}_{\ref{diag:qgqbar}}^{(1;2)}]^2  \mathrm{Re}[1-\mathcal{S}_{012}]  - \frac{e^{-\vert\xt_{2;\ref{diag:qgqbar}}\vert^2/(\vert\xt_{01}\vert^2 e^{\gamma_E})}}{\vert\xt_{2;\ref{diag:qgqbar}}\vert^2} \biggl [K_0 \left (\vert \xt_{01}\vert \sqrt{\overline{Q}^2_{\ref{diag:qgqbar}} + m^2} \right )\biggr ]^2   \mathrm{Re}[1-\mathcal{S}_{01}]  \biggr \} \\
& + (k^+_0)^2\biggl [2k^+_1(k^+_1 + k^+_2) + (k^+_2)^2 \biggr ]  \\
& \hspace{1.5cm} \times \biggl \{   \frac{\vert\xt_{2;\ref{diag:qqbarg}}\vert^2}{64}[\mathcal{G}_{\ref{diag:qqbarg}}^{(1;2)}]^2  \mathrm{Re}[1-\mathcal{S}_{012}]  - \frac{e^{-\vert\xt_{2;\ref{diag:qqbarg}}\vert^2/(\vert\xt_{01}\vert^2 e^{\gamma_E})}}{\vert\xt_{2;\ref{diag:qqbarg}}\vert^2} \biggl [K_0 \left (\vert \xt_{01}\vert \sqrt{\overline{Q}^2_{\ref{diag:qqbarg}} + m^2} \right )\biggr ]^2   \mathrm{Re}[1-\mathcal{S}_{01}]  \biggr \} \\
& - \frac{k^+_0k^+_1}{32} \biggl [k^+_1(k^+_0 + k^+_2) + k^+_0(k^+_1 + k^+_2) \biggr ] (\xt_{2;\ref{diag:qgqbar}}\cdot \xt_{2;\ref{diag:qqbarg}}) [\mathcal{G}_{\ref{diag:qgqbar}}^{(1;2)}][\mathcal{G}_{\ref{diag:qqbarg}}^{(1;2)}]\mathrm{Re}[1-\mathcal{S}_{012}]\\
& + \frac{m^2}{16}(k^+_2)^4 \Biggl [\frac{k^+_1}{(k^+_0 + k^+_2)} [\mathcal{G}_{\ref{diag:qgqbar}}^{(1;1)}]  - \frac{k^+_0}{(k^+_1 + k^+_2)}[\mathcal{G}_{\ref{diag:qqbarg}}^{(1;1)}] \Biggr ]^2
\mathrm{Re}[1-\mathcal{S}_{012}]\Biggr \}.
\end{split}
\end{equation}

\end{widetext}

\bibliography{spires}
\bibliographystyle{JHEP-2modlong}

\end{document}